\documentclass[12pt]{article}  
\usepackage{graphicx}
\usepackage{amssymb}
\input epsf
\parskip=\medskipamount
\def\CaE{{\cal E}}

\def\al{\alpha}

\def\la{\lambda}  
\def\k{\kappa}  
\def\kp{\kappa}  
   
\def\om{\omega}   \def\Om{\Omega}
   
\def\IC{\relax{\rm l\kern-.50 em C}}
\def\IE{\relax{\rm l\kern-.12 em E}}
\def\IK{\relax{\rm l\kern-.18 em K}}
\def\IL{\relax{\rm I\kern-.18 em L}}
\def\IN{\relax{\rm I\kern-.18 em N}}
\def\IR{\relax{\rm I\kern-.18 em R}}

\font\tenfrak=eufm10  \font\sevenfrak=eufm7  \font\fivefrak=eufm5
\newfam\frakfam
\textfont\frakfam=\tenfrak
\scriptfont\frakfam=\sevenfrak
\scriptscriptfont\frakfam=\fivefrak

\newtheorem{proposicion}{Proposition}

\def\wh{\widehat}
\def\wt{\widetilde}
\def\frac#1#2{{#1\over #2}}

\def\ptos{\leaders\hbox to 2mm{\hfil{.}\hfil}\hfill}

\begin{document}

\title{ A Quantum Quasi-Harmonic Nonlinear  Oscillator \\  with an Isotonic Term }

\author{ Manuel F. Ra\~nada$^{a)}$  \\ [3pt]
 {\sl Dep. de F\'{\i}sica Te\'orica and IUMA } \\
 {\sl Universidad de Zaragoza, 50009 Zaragoza, Spain}     }
\maketitle

\begin{abstract}
  The properties of a nonlinear oscillator with an additional term $k_g/x^2$, 
characterizing the isotonic oscillator,  are studied.
The nonlinearity affects to both the kinetic term and the potential
and combines two nonlinearities associated to two parameters, 
$\kappa$ and $k_g$,  in such a way that for $\kappa=0$ all
the characteristics of of the standard isotonic system are recovered.
The first part is devoted to the classical system and the second part to the quantum system. 
This is a problem of quantization of a system with position-dependent mass of the form $m(x)=1/(1 - {\kappa} x^2)$, with a $\kappa$-dependent non-polynomial rational potential and with an additional isotonic term. 
The Schr\"odinger equation is exactly solved and the $(\kappa,k_g)$-dependent wave functions and bound state energies are explicitly obtained for both $\kappa<0$ and $\kappa>0$. 
\end{abstract}

\begin{quote}


\end{quote}
{\vfill}

\footnoterule
{\noindent\small
$^{a)}${\it E-mail address:} {mfran@unizar.es}
\newpage


\section{Introduction }

 The following potential
\begin{equation}
 V_{\rm Isot}(x) = V_0(x) + V_g(x) = (\frac{1}{2})\,\om^2\, x^2
 +  (\frac{1}{2})\,\frac{k_g}{x^2} \,,{\quad} g>0\,,
\end{equation}
representing an harmonic oscillator with an additional term similar to a centripetal barrier, is known as the isotonic oscillator (or singular harmonic oscillator). 
It is important because, although nonlinear, is endowed with properties closely related with
those of the harmonic oscillator \cite{WeiJo79}--\cite{GranQ13}. 
At the classical level, the Euler-Lagrange equation, that is given by
\begin{equation}
 \ddot{x} + \om^2\,x - \frac{k_g}{x^3} = 0 \,,  \label{EqPinney}
\end{equation}
is a particular case of the Pinney-Ermakov equation \cite{Pin50}.
It can be exactly solved and the solution is given by
$$
 x = \frac{1}{\om\,A}\,\sqrt{(\om^2A^4 - k_g)\,\sin^2(\om t + \phi) + k_g\,}\,.
$$
showing explicitly the periodicity of the solutions.  At the
quantum level the Schr\"odinger equation, that takes the form
$$
 - \frac{\hbar^2}{2 m}\,\frac{d^2\Psi}{dx^2}
 +  \frac{1}{2}\,\Bigl[m\om^2\,x^2 +  \frac{k_g}{x^2} \Bigr]
 \,\Psi = E\,\Psi   \,, 
$$
can be reduced (introducing the appropriate changes) to a confluent hypergeometric function in such a way that the energy eigenfunctions $\Psi_n$ are characterized by energies 
$$
  E_n =  \Bigl((2 n+1)+ \frac{1}{2} + g\Bigr)(\hbar\omega) \,,{\quad} n=0,1,2,\dots
$$
The following two points summarize the main characteristis of this system. 
\begin{itemize}
\item     At the classical level, the Lagrange equation is a nonlinear but exactly solvable equation and the system is isochronous, that is, the period of the oscillations  is independent of the amplitude (or of the energy). 
\item     At the quantum level, the system is  exactly solvable and the energy spectrum is equidistant. Nevertheless, the height $\Delta E =E_{n+1} - E_n$ of the energy steps is twice that of the simple harmonic oscillator. In fact, it seems as if the new additional term $k_g/x^2$ causes the vanishing of half of the levels of the original linear system.

\end{itemize}

In addition to these properties we can add that the two dimensional version of this oscillator, that is known as the Smorodinsky-Winternitz system, is separable in two different systems of coordinates and it is therefore superintegrable with quadratic constants of motion.

On the other hand  Mathews and Lakshmanan studied in 1974 \cite{MaLak74},\cite{LakRa03}, the differential equation
\begin{equation}
  (1 +\la x^2)\,\ddot{x} - (\la x)\,\dot{x}^2 + \al^2\,x  = 0
 \,,{\quad}\la>0\,,                 \label{EqMaL}
\end{equation}
and they proved that the general solution is of the form 
$$
  x  = A \sin(\om\,t + \phi) \,,
$$
with the following additional restriction linking frequency and
amplitude
$$
  \om^2  = \frac{\al^2}{1 + \la\,A^2} \,.
$$
The equation (\ref{EqMaL}) is therefore an interesting example of a
system with nonlinear oscillations with a frequency (or period)
showing amplitude dependence. It is a lagrangian equation with Lagrangian 
$$
  L(x,v_x;\lambda) = \frac{1}{2}\,\Bigl(\frac{v_x^2}{1 + \lambda\,x^2} 
\Bigr) -     \frac{1}{2}\,\Bigl(\frac{\al^2\,x^2}{1 + \lambda\,x^2} \Bigr)
$$
As a quantum system, the Schr\"odinger equation involving
the potential $x^2/(1 + \lambda\, x^2)$ has been studied by different authors using different approaches \cite{Fle81}--\cite{BagDas13}. 
In addition to the nonpolynomial character of the potential this system is also interesting because it is a particular case of a system with a position-dependent mass (see \cite{Le95}--\cite{LimaV12} and references therein). 
This poses an important problem since some authors have proposed some different ways of carrying  out the process of quantization. 

We have studied this quantum nonlinear system in \cite{CaRS04}--\cite{CaRS07a} using as a method for quantization the idea that the quantum Hamiltonian, and also other related operators, must be self-adjoint but  in a Hilbert space determined by a measure $d\mu_{\lambda}$ that depends on the parameter $\lambda$. 
Now, we continue with the study of this particular nonlinear  oscillator. 
We have just seen that the quantum isotonic oscillator is exactly solvable; therefore it seems reasonable to study the quantum isotonic version of this nonlinear system. The main objective of this article is the study of the quantum nonlinear system but we have considered appropriate to first present the main characteristics of the classical system.

  The plan of the article is as follows:
Sec. 2 is devoted to study of the nonlinear oscillator with an Isotonic term $k_g/x^2$ from the view point of a classical dynamical system (this section is related with some questions studied in \cite{CaRSS04}--\cite{CaRS05}) and then in Sec. 3 we study first the quantization of the system (this problem is related with the quantization studied in  \cite{CaRS04} and  \cite{CaRS07a}) and then we consider the Schr\"odinger equation.  
It is proven  that this system  is Schr\"odinger solvable and then the wave functions $\Psi_n$ and the energies $E_n$ of the bound states are explicitly obtained.
Finally in Sec. 4 we make some final comments.

\section{Nonlinear oscillator with an Isotonic term }

\subsection{Isotonic Harmonic Oscillator }

 Let us now consider the following $\kappa$-dependent Lagrangian
\begin{equation}
  L(x,v_x;\kp,k_g) = \frac{1}{2}\,\Bigl(\frac{v_x^2}{1 - \kp\,x^2} 
\Bigr) -  V_{\kp,g}(x)
\,,{\quad} V_{\kp,g}(x) = \frac{1}{2}\,\Bigl(\frac{\al^2\,x^2}{1 - \kp\,x^2} \Bigr)
+ \frac{1}{2}\,\frac{k_g}{x^2}\,,\ k_g>0\,, \label{Lagxlak}
\end{equation}
that corresponds to the nonlinear oscillator of Mathews and Lakshmanan
with an additional term of the form $k_g/x^2$.
The parameter $\kappa$ can take both positive and negative values; nevertheless
as for $\kappa>0$, the Lagrangian (and the associated  dynamics)
will have a singularity at $1 -\,\kp\,x^2=0$, we restrict the study of the
dynamics to the interior of the interval $x^2<1/|\kp|$ that is the region
in which kinetic term is positive definite.
Figure I and II show the form of the potential $V_{\kp,g}(x)$ for 
several values of $\kappa$ ($\kappa>0$ in Figure I and $\kappa<0$ in Figure II).

The Euler-Lagrange equation is
\begin{equation}
 \frac{d^2x}{dt^2} + \frac{\kp\,x}{1 - \kp\,x^2}\,\Bigl(\frac{d x}{dt}\Bigr)^2
 +  \frac{\al^2 x}{1 - \kp\,x^2}
 - k_g\, \Bigl(\frac{1 - \kp\,x^2}{x^3}\Bigr) = 0 \,,
\label{Eqkappag}\end{equation}
in such a way that for $\kappa\to 0$ we recover the Pinney equation (\ref{EqPinney}).

This  equation combines two nonlinearities associated to the two parameters, $k_g$ and $\kappa$. It looks certainly difficult to be solved but, nevertheless, the general solution can be obtained by assuming  for the function $x(t)$ certain  particular expressions depending of some undetermined coefficients.

\subsection{Positive  $\kappa>0$ case }  

If $\kappa>0$ then all the solutions are periodic solutions. 

Let us suppose, as an ansatz, that the general solution of the equation $\kappa> 0$ is quite similar to the solution of the $\kappa=0$ equation, that is,  
\begin{equation}
  x = \Bigl(\frac{1}{\om\,A}\Bigr)\sqrt{(\om^2A^4 - k_g)\,\sin^2(\om t  + \phi)   + k_g\,} \,,
\label{x1t}
\end{equation}
then the equation (\ref{Eqkappag}) leads to the following algebraic equation  
$$
  \kp\,\om^2\,A^4 + (\al^2 - \om^2 -  k_g \kp^2)\,A^2 +  k_g \kp = 0 \,.
$$
Therefore the function (\ref{x1t}) is in fact a solution of
(\ref{Eqkappag})  but where $\om$, that determines the angular
frequency of the motion, is $(\kp,k_g)$-related with the coefficient $\al$
of the potential (which represents the frequency of the $(\kp=0,k_g=0)$
harmonic oscillator) by
$$
 \om^2 = \frac{\al^2}{(1 - \kp\,A^2)} +  \frac{k_g \kp}{A^2} \,.
$$
The solution (\ref{x1t}) oscillates between $x_-= \sqrt{k_g}/(w A)$ and $x_+=A$.

The energy is given by
$$
 E =  \frac{1}{2}\,\al^2\,\Bigl(\frac{A^2}{\,1 - \kp\,A^2\,}\Bigr)
 + \frac{1}{2}\,\frac{k_g}{A^2}  \,.
 $$
so that 
$$
   \om^2 = \al^2 + 2 \kp E \,.
$$
Note that the coefficient $(1 - \kp\,A^2)$ is positive even for
$\kappa>0$ since in that case the amplitude $A$ must satisfy
$A^2<1/\kp$. 
Notice also that when $k_g\to 0$ these expressions reduce to
$$
  x = A\,\sin(\om t + \phi)  \,,{\quad}  \al^2 = (1 - \kp\,A^2)\,\om^2    
  \,,{\quad}
  E = \bigl(\frac{1}{2}\bigr)\,\al^2\,\Bigl(\frac{A^2}{\,1 - \kp\,A^2\,}\Bigr) \,,
$$
which are just the relations obtained in \cite{MaLak74}, \cite{CaRSS04}.

  We summarize: 
the solution $x(t)$ of the dynamics depends of the three coefficients $A$, $\phi$ and $\om$; 
two coefficients, $A$ and $\phi$, remain arbitrary but $\om$ 
becomes a $(\kp,k_g)$-dependent function of the amplitude $A$. 
In this case, as the parameter $\kappa$ is positive $\kappa>0$, we have  $\om>\al$.

\subsection{Negative  $\kappa<0$ case }  

If $\kappa<0$ then there are two possible behaviours : bounded motions (for energies lower than a certain value $E_b$) and unbounded motions (energies greater than that value).

\subsubsection{Bounded $\kappa<0$ motions }  

If $\kappa<0$ then the system also admits periodic solutions of the form
\begin{equation}
  x = \Bigl(\frac{1}{\om\,A}\Bigr)\sqrt{(\om^2A^4 - k_g)\,\sin^2(\om t  + \phi)   + k_g\,} \,,
\label{x2t}
\end{equation}
but now $\omega$, $\alpha$, and the energy $E$ are related by 
$$
   \om^2 = \al^2 - 2 |\kp| E 
$$
so the energy $E$  must satisfy the inequality
$$
 E < E_b   {\quad}{\rm with}{\quad} E_b = \frac{\al^2}{2 |\kp|}  \,.
$$
Thus the allowed values of $E$ are bounded by $E_b$ with the value of $\om$ decreasing when the energy $E$ approaches to this upper value.

We can summarize these nonlinear periodic oscillations as follows. 
\begin{itemize}
\item[(a)]     If the parameter $\kappa$ is negative $\kappa<0$, then $\om<\al$.
\item[(b)]     If the parameter $\kappa$ is positive $\kappa>0$, then $\om>\al$.
\end{itemize}

\subsubsection{Unbounded motions}

If we the parameter $\kappa$ is negative and the energy is greater than $E_b$, that is $\kappa<0$ and $E_b<E$, then the solution of the dynamics is given by 
\begin{equation}
  x = \Bigl(\frac{1}{\Om\,A}\Bigr)\sqrt{(\Om^2A^4 + k_g)\,\sinh^2(\Om t 
+ \phi) + k_g\,} \,,
\label{x2t}\end{equation}
with the additional constraint 
$$
   \kp\,\Om^2\,A^4 + (\al^2 + \Om^2 - k_g \kp^2 )\,A^2  - k_g \kp = 0 \,. 
$$
Solving this equation we obtain the following expresion for  $\Om$ (hyperbolic frequency) as a function of $\al$
$$
 \Om^2 = \frac{\al^2}{|\kp|\,A^2 -1} - \frac{k_g |\kp|}{A^2}   \,. 
$$
The energy, that now we denote by $E_h$,  is given by
$$
 E_h = \frac{1}{2}\,\al^2\,\Bigl(\frac{A^2}{|\kp|\,A^2 - 1}\Bigr) - \frac{1}{2}\,\frac{k_g}{A^2}  \,.  
$$
Using the expressions of $\Om^2$ and $E_h$ we arrive to
$$
  \Om^2 = 2 |\kp| E_h - \al^2  
$$
and, since $\Om^2>0$, we conclude that
$$
 E_h > \frac{\al^2}{2 |\kp| } \,.
$$
Notice also that when $k_g\to 0$ these expressions reduce to
$$
 x = A\,\sinh(\Om t + \phi)  \,,{\quad}
 \al^2 = (|\kp|\,A^2 -1 )\,\Om^2   \,,{\quad}
 E = \frac{1}{2}\,\al^2\,\Bigl(\frac{A^2}{|\kp|\,A^2 - 1}\Bigr) \,, 
$$
and coincide with the values obtained for the $\kappa$-case in \cite{CaRSS04}.

\subsubsection{Border unbounded motions}

The Lagrange equation (\ref{Eqkappag}) also admits the following algebraic function 
\begin{equation}
  x = \sqrt{A\,t^2 + B\, t+ C\,} \,, 
\label{x3t}\end{equation}
with $A$ and $C$ taken the following values
$$
 A = \frac{k_g \kp^2- \al^2}{\kp}   \,,\qquad
 C   = -\,\frac{(B^2 + 4 k_g)\kp}{4(\al^2 - k_g \kp^2)}  
  \,,\qquad (\kp<0)\,, 
$$
as solution. This very particular solution represents the border between the trigonometric (periodic) solutions and the hyperbolic (unbounded) solutions 
(they play a rather similar role to the parabolic trajectories in the Kepler problem).
The associated energy is just $E = E_b$ as was to be expected.

\section{Quantum nonlinear oscillator with an Isotonic term }

\subsection{Quantization }

The momentum $p_x$ is given by $p_x=v_x/(1 - \kp\,x^2)$ so that the $(\kp,g)$-dependent Hamiltonian of the system is 
\begin{equation}
  H(x,p_x;\kp,g) = \frac{1}{2m}\,\Bigl(1 - \kp\,x^2\Bigr)\,p_x^2 
  + \frac{1}{2}\,m\,\al^2\, \Bigl(\frac{x^2}{1 - \kp\,x^2}\Bigr)  
  + \frac{1}{2}\,\frac{k_g}{x^2}                    \label{HClasico}  \,. 
\end{equation}
It is clear that it is a PDM system, that is a sytem with a position dependent mass  \cite{Le95}--\cite{LimaV12}. 
 The important point is that if the mass $m$ becomes a spatial function,
$m=m(x)$, then the quantum version of the mass no longer commutes
with the momentum.    Therefore, different forms of presenting the
kinetic term $T$ in the Hamiltonian $H$, as for example 
$$
 T = \frac{1}{4}\,\Bigl[\frac{1}{m(x)}\,p^2 + p^2\,\frac{1}{m(x)}\Bigr]    \,,{\quad}
 T = \frac{1}{2}\,\Bigl[p\,\frac{1}{m(x)}\,p \Bigr]    \,,{\quad}
 T = \frac{1}{2}\,\Bigl[\frac{1}{\sqrt{m(x)}}\,p^2\,\frac{1}{\sqrt{m(x)}}\Bigr]  \,,
$$
are equivalent at the classical level but they lead to different and 
nonequivalent Schr\"odinger equations.

In this case the construction of the appropriate quantum  Hamiltonian rests on the idea that, because of the $x$-dependence of the kinetic term, the Hilbert space of the quantum system must be, not the standard space $L^2(\IR,dx)$, but the space $L_\kp^2(d\mu_\kp)$  where $d\mu_\kp$ denotes the following $\kappa$-dependent measure  
$$
 d\mu_\kp = \frac{dx}{\sqrt{1 - \kp\,x^2}}
$$ 
and the particular form of the Hilbert space $L_\kp^2(d\mu_\kp)$ depends on $\kappa$ as follows 

\begin{itemize}

\item[(a)]     Negative $\kappa<0$ case. The space $L_\kp^2(d\mu_\kp)$ can be identified with $L^2(\IR^+,d\mu_\kp)$, $\IR^+=[0,\infty)$. 
\item[(b)]    Positive $\kappa>0$ case. The space $L_\kp^2(d\mu_\kp)$ can be identified with $L_0^2(I_\kp,d\mu_\kp)$ where $I_\kp$ denotes the interval $[0,b_\kp]$, $b_\kp=1/\sqrt{\kp}$,  and the subscript means that the functions must vanish at the  endpoints. 

\end{itemize}

The quantum Hamiltonian $\widehat{H}(\kp)$  must be self-adjoint in the space $L_\kp^2(d\mu_\kp)$. 
Now, we note that the Hamiltonian can be rewritten as  
$$
  H(\kp) = \bigl(\frac{1}{2m}\bigr)\,P_x^2
  +  \frac{1}{2}\,m\,\al^2\, \Bigl(\frac{x^2}{1 - \kp\,x^2}\Bigr) 
  +  \frac{1}{2}\,\frac{k_g}{x^2} \,,
  {\quad} P_x = \sqrt{1 - \kp\,x^2}\,p_x  \,. 
$$
Thus, for obtaining the expression  of the operator  $\widehat{H}(\kp)$ we first consider  the  operator $\widehat{P_x}(\kp)$,  representing the quantum version of of the Noether momentum  $P_x(\kp)$.

\begin{proposicion} The operator $\wh{P_x}$ 
$$
 \wh{P_x} = -\,i\,\hbar\,\,\sqrt{1 - \kp\,x^2}\,\,\,d/dx \,,
$$ 
is self-adjoint   in the space $L_\kp^2(d\mu_\kp)$. 
\end{proposicion}
 
Therefore, the transition from the classical system to the quantum one is given by following correspondence 
$$
 P_x \,\mapsto\, \wh{P_x} = -\,i\,\hbar\,\sqrt{1 - \kp\,x^2}\,\frac{d}{dx}  \,,
$$
so that
$$
 (1 - \kp\,x^2)\,p_x^2 \,\mapsto\, -\,\hbar^2\,
 \Bigl(\sqrt{1 - \kp\,x^2}\,\frac{d}{dx}\Bigr)
 \Bigl(\sqrt{1 - \kp\,x^2}\,\frac{d}{dx}\Bigr) \,,
$$
in such a way that the quantum version of the Hamiltonian
(\ref{HClasico}) becomes
$$
 \widehat{H} = - \frac{\hbar^2}{2m}\,(1 - \kp\,x^2)\,\frac{d^2}{dx^2}
 + \bigl(\frac{\hbar^2}{2m}\bigr)\,\kp\,x\,\frac{d}{dx}
 + \bigl(\frac{1}{2}\bigr)\,\al^2 \Bigl(\frac{x^2}{1 - \kp\,x^2}\Bigr)+  \frac{1}{2}\,\frac{k_g}{x^2} \,.
$$

\subsection{Schr\"odinger equation }
The Schr\"odinger equation, that is given by 
$$
 \Bigl[ - \frac{\hbar^2}{2 m}\,(1 - \kp\,x^2)\,\frac{d^2}{dx^2}
 + \bigl(\frac{\hbar^2}{2m}\bigr)\,\kp\,x\,\frac{d}{dx}
 + \bigl(\frac{1}{2}\bigr)\,m\al^2 \Bigl(\frac{x^2}{1 - \kp\,x^2}\Bigr) 
 +  \frac{1}{2}\,\frac{k_g}{x^2} \Bigr]
 \,\Psi = E\,\Psi   \,, 
$$
can be simplified by introducing adimensional variables 
$$
 \rho^2 = \mu^2\,x^2  \,,{\quad}
 \kp = \mu^2\,\kp'   \,,{\quad}
  E = (\hbar\,\al)\,\CaE  \,,{\quad}
 k_g = \frac{\hbar^2}{m} \,g(g+1) \,,{\quad}
 \mu^2 = \frac{m\,\al}{\hbar}    \,,
$$
so that it  becomes
\begin{equation}
 (1 - \kp'\,\rho^2)\, \frac{d^2\Psi}{d\rho^2} - \kp' \,\rho\,\frac{d\Psi}{d\rho} 
 - (1 - \kp')\Bigl(\frac{\rho^2}{1 - \kp'\,\rho^2}\Bigr)\,\Psi- \frac{g(g+1)}{\rho^2} + (2\,\CaE)\,\Psi  = 0  \,.
\label{EqPsi(yLa)}
\end{equation}
Next we proceed in several steps. 
\begin{itemize}
\item   Step 1. 
We assume the following factorization for $ \Psi(\rho,\kp',g)$  
$$
  \Psi(\rho,\kp',g) = h(\rho,\kp',g)\,(1 - \kp'\,\rho^2)^{\,1/(2\kp')}  \,, 
$$
where the second factor is an algebraic function that satisfies the limit 
$$
 \lim{}_{\kp'{\to}0}\,(1 - \kp'\,\rho^2)^{\,1/(2\kp')}  =  e^{-\,(1/2)\,\rho^2}\,.
$$
Then the new function $h$ must satisfy the differential
equation
$$
  (1 - \kp'\,\rho^2) h'' - (2+\kp') \rho\, h'  -\frac{g(g+1)}{\rho^2}\,h + (2 \CaE -1) h = 0 \,,{\qquad}   h=h(\rho,\kp',g)\,. 
$$
  If $\kappa=0$ then we obtain
$$
   h'' - 2 \rho h'  -\frac{g(g+1)}{\rho^2}\,h + (2 \CaE -1) h = 0 \,,{\qquad} h=h(\rho,g)
$$
\item   Step 2. 
Now we introduce a  factorization for  $h(\rho,\kp)$  
$$
 h = \rho^{(g+1)} \,w(\rho)
$$
Then the new function $w(\rho,\kp)$ must satisfy the differential
$$
  (1 - \kp'\,\rho^2) w'' + \Bigl[2\,  \frac{g+1}{\rho} - (2 + 3\kp'\, + 2 g\,\kp')\,\rho\Bigr]\,w'     + \Bigl[2 \Bigl(  \CaE -  \bigr( g + \frac{3}{2}\bigr)\,\Bigr) -\kp'\,(1+g)^2\,\Bigr] \,w = 0 \,. 
$$
\item   Step 3.     A new change of variable  
$$
  \rho\ \to\ z = \rho^2
$$
leads to the following equation 
$$
 z\, (1 - \kp'\,z)\,\frac{d^2w}{dz^2} +  
  \Bigl[ g +  \frac{3}{2}  - (1 + 2\kp'\, +  g\,\kp') \,z\Bigr]\,\frac{dw}{dz}  
 + \Bigl[ \frac{1}{2}\, \Bigl(  \CaE -  g - \frac{3}{2}\bigr) -\frac{\kp'}{4}\,(1+g)^2\,\Bigr] \,w = 0 \,. 
$$
\item   Step 4.   Finally, introducing $t$ defined as  $t = \kp'\,z$, $\kp' \ne 0$,  we obtain 
 $$
  t\, (1 - t)\,\frac{d^2w}{dt^2}  +  
  \Bigl[\,g +  \frac{3}{2}-(\frac{1}{k'} +2+g)\,t \Bigr]\,\frac{d\,w}{dt}  + \Bigl[ \frac{1}{2\,\kp'}\, \Bigl(  \CaE -  g - \frac{3}{2}\bigr) -\frac{1}{4}\,(1+g)^2\,\Bigr] \,w = 0 
  $$that is a hipergeometric equation 
 $$
  t\, (1 - t)\,\frac{d^2w}{dt^2}  +  \Bigl[\,c-(1+a_\kp+b_\kp)\,t \,\Bigr]\,\frac{d\,w}{dt} - a_\kp b_\kp \,w = 0 
$$
with $c = g + 3/2$ and 
$$
a_\kp = \frac{1}{2}\Bigl(1 + g + \frac{1+ \sqrt{1 - \kp' + 2\,\kp'\, \CaE\, }}{\kp'}\, \Bigr) 
\,,{\quad}
b_\kp = \frac{1}{2}\Bigl(1 + g + \frac{1- \sqrt{1 - \kp' + 2\,\kp'\, \CaE\, }}{\kp'}\, \Bigr) \,. 
$$
\end{itemize}
As it is well known, if the coefficient  $a_\kp$ or $b_\kp$  is a nonnegative integer,  $a_\kp=-n$  or  $b_\kp=-n$, then the hipergeometric series ${}_2F_1(a_\kp,b_\kp\,;c\,;t)$ has only a finite number of terms and in fat  it becomes a polynomial of degree $n$. 

The polynomials solutions ${\cal P}_n(t)$ of the above equation are given by  
$$
   {\cal P}_n(t) = {}_2F_1(-n,\,b_{\kp n} \,;\, g +  \frac{3}{2} \,;\,t)
$$
with $b_{\kp n}$ representing the value of $b_\kp$ when $a_\kp=-n$
$$
  \ b_{\kp n} = 2(n + 1 + g + 1/\kp') \,. 
$$

Finally, we note that the last change (step 4)  is  necessary because of the presence of the parameter $\kappa$. In the more simple $\kappa=0$ case the Schr\"odinger equation leads to a confluent equation; the presence of $\kappa\ne 0$ transforms the confluent equation into a more general hypergeometric equation. 

\subsection{$\kappa$-dependent Sturm-Liouville problems and orthogonality}

In what follows, and for easy of notation, we just write $\kappa$ instead of $\kp'$. 

The $\kappa$-dependent differential equation for $w$ 
$$
 a_0 w'' + a_1 w' + a_2 w = 0 \,,
$$
$$
 a_0 =  1 - \kp\,\rho^2  \,,{\quad} 
 a_1 =  2 \frac{g+1}{\rho} - (2 + 3\kp'\, + 2 g\,\kp)\,\rho  \,,{\quad}
 a_2 =   \Bigr(\CaE -  g - \frac{3}{2}\,\Bigr) -\kp\,(1+g)^2  \,,{\quad}
$$
is not self-adjoint but it  can be reduced to self-adjoint form by making use of the following  factor
$$
 \la(\rho) = (\frac{1}{a_0})\,e^{{\int}(a_1/a_0)\,d\rho}
 = \rho^{2(g+1)}\,(1 - \kp\,\rho^2)^{1/\kp-1/2}  \,, 
$$
so that the equation becomes 
$$
 \frac{d}{dx}\Bigl[\,p(\rho,\kp)\,\frac{dh}{d\rho}\,\Bigr]
 + (2 e -1)\,r(\rho,\kp)\,h = 0 \,,
$$
where the  $p=p(\rho,\kp)$ and $r=r(\rho,\kp)$ are given by
$$
  p(\rho,\kp)  = \rho^{2(g+1)}\,\sqrt{1 - \kp\,\rho^2}\,(1 - \kp\,\rho^2)^{1/\kp}\,, {\qquad}
  r(\rho,\kp) = a_2\,\rho^{2(g+1)}\,(1 - \kp\,\rho^2)^{1/\kp-1/2}\,.
$$
This equation, together with appropriate conditions for the behaviour of
the solutions at the end points, constitute a  Sturm-Liouville problem. 
As the boundary conditions are in fact different according to the sign of the parameter $\kappa$ we arrive to, no just one, but two different 
Sturm-Liouville problems:

\begin{itemize}

\item[(a)]   Negative $\kappa<0$ case  
 
The variable $\rho$ is defined in the half real line $\IR^+=[0,\infty)$ and, therefore, the S-L problem is singular.   The solutions $w(\rho,\kp)$ must be well defined in all $\IR^+$ and the boundary conditions prescribe that the behaviour of these functions when $\rho\to\,\infty$ must be such that their norms, determined with respect to the weight function $r(\rho)$, be finite.

The solutions of the problem are the $\kappa$-dependent polynomials ${\cal P}_m$, $m=0,1,2,\dots$

 \item[(b)]    Positive $\kappa>0$ case 
 
The range of the  variable $\rho$ is limited by the restriction $\rho^2<1/\kp$. 
and the problem is  defined in the bounded interval $[0,b_\kp]$ with $b_\kp=1/\sqrt{\kp}$.   It is singular because the function $p(\rho,\kp)$ vanishes in the two end points $\rho_1=0$ and $\rho_2=b_\kp$. In this case the solutions $w(\rho,\kp)$ of the problem must be bounded functions at $\rho_1=0$ and $\rho_2=b_\kp$ (if $w$ is bounded then the wave function $\Psi$ vanishes).

We obtain the above mentioned polynomial solutions.

\end{itemize}

\begin{proposicion}
 The eigenfunctions of the S-L problem ($\kappa<0$ and $\kappa>0$) are orthogonal with respect to the function $r= \rho^{2(g+1)}\,(1 - \kp\,\rho^2)^{1/\kp-1/2}$. 
\end{proposicion} 
{\it Proof:} This statement is just a consequence of the
properties of the Sturm-Liouville problems.

Because of this the polynomial solutions ${\cal P}_m$,
$m=0,1,2,\dots$,  satisfy
$$   (a)  {\quad} 
  \int_{0}^{\infty} \Bigl({\cal P}_m(\rho,\kp)\,{\cal P}_n(\rho,\kp)\Bigr)
  \frac{(1 - \kp\,\rho^2)^{1/\kp}}{\sqrt{1 - \kp\,\rho^2} } \,\rho^{2(g+1)}\,\,d\rho = 0  \,,{\quad} m\,\ne\,n \,,{\quad} \kp<0\,, 
$$
and
$$   (b) {\quad} 
  \int_{0}^{1/\sqrt{\kp}} \Bigl({\cal P}_m(\rho,\kp)\,{\cal P}_n(\rho,\kp)\Bigr)
  \frac{(1 - \kp\,\rho^2)^{1/\kp}}{\sqrt{1 - \kp\,\rho^2} }\,\rho^{2(g+1)}\,\,d\rho = 0
 \,,{\quad} m\,\ne\,n \,,{\quad} \kp>0\,. 
$$

If we define the  functions  $\Psi_n$ by
$$
 \Psi_n(\rho,g) = {\cal P}_n(\rho,g)\,\rho^{(g+1)}\,(1 - \kp\,x^2)^{1/2\kp}
 \,,{\quad}  n=0,1,2,\dots
$$
then the above statement admits the following alternative form:
{\sl The $\kappa$-dependent  functions $\Psi_n(\rho,g)$ are
orthogonal with respect to the weight function  $\wt{r}=1/\sqrt{1
- \kp\,\rho^2}$}.  That is 
$$   (a)  {\quad} 
 \int_{0}^{\infty} \Psi_m(\rho,\kp)\,\Psi_n(\rho,\kp)\,\wt{r}(\rho,\kp) \,dx
 = \int_{0}^{\infty}  \Psi_m(\rho,\kp)\,\Psi_n(\rho,\kp)\,d\mu_\kp  = 0
\,,{\quad} m\,\ne\,n \,,{\quad} \kp<0\,,  
$$
and
$$   (b)  {\quad} 
 \int_{0}^{1/\sqrt{\kp}} \Psi_m(\rho,\kp)\,\Psi_n(\rho,\kp)\,\wt{r}(\rho,\kp) \,dx
 = \int_{0}^{1/\sqrt{\kp}} \Psi_m(\rho,\kp)\,\Psi_n(\rho,\kp)\,d\mu_\kp = 0
 \,,{\quad} m\,\ne\,n \,,{\quad} \kp>0\,,
$$
where we recall that $d\mu_\kp=\wt{r}(\rho,\kp) \,dx$ represents the $\kappa$-dependent mesure introduced in the quantization of the momentum and the Hamiltonian. 

We close this section pointing out the importance of this result, the orthogonality relations associated to the Sturm-Liouville problem are with respect the measure $d\mu_\kp$ and therefore they  are consistent with the Hilbert space structure introduced for the quantization of the system as a position dependent mass (PDM) system.

\subsection{Wave functions and energy levels }

We have arrived to the following situation 
 
\begin{itemize}
\item   Bound state  wave functions
$$
 \Psi_n = N_n\,  (\mu\,x)^{\,g+1}\,(1 - \kp\,\mu^2x^2)^{\,1/(2\kp)}\, {\cal P}_n(x)  \,,{\quad}
 {\cal P}_n(x) = {}_2F_1(-n,b_\kp\,;\, g+3/2\,;\, \kp\,\mu^2x^2) \,. 
$$
\item  Bound state energy eigenvalues $\CaE_n$ 
$$
  \CaE_n =  \Bigr(m+ \frac{1}{2} +  g\Bigr) + \frac{1}{2} \kp\,(m + g )^2  
  \,,{\quad}    m= 2 n + 1  \,. 
$$

\end{itemize}

 Nevertheless the  $\kappa<0$ situation deserves be studied with more detail.
We recall that the equation (and the solutions) is defined on $\IR^+=[0,\infty)$  and hence it is necessary to take  into account the problem of the convergence at the   infinity.  In fact, it is necessary that the following integral be convergent 
$$    
  \int_{0}^{\infty} \Bigl({\cal P}_n(\rho,\kp)\Bigr)^2
  \frac{\rho^{2(g+1)}}{(1 + |\kp|\,\rho^2)^{1/|\kp|}\sqrt{1 + |\kp|\,\rho^2}} \,\,d\rho < \infty
  \,,{\quad} \kp<0\,, 
$$    
and, as for large values of $x$ the powers of the dominant terms in the numerator and the denominator are $4 n + 2(g+1)$ (${\cal P}_n$ has only even powers) and $1 + 2/|\kp|$, respectively, we arrive to a certain condition to be satisfied by $n$. In fact, given a certain value of $|\kp|$, then the admissible functions ${\cal P}_n$ are those  associated to integer values of $n$ satisfying the condition 
$$
 n < N_\kp = \frac{1}{2|\kp|}\Bigl( 1 - (g+1)|\kp|\Bigr)  \,. 
$$
Thus, when $|\kp|$ (with $\kappa<0$) increases its value then the number of eigenstates decreases and for great values of $|\kp|$ the system only admits the fundamental level as stationary solution. 

An alternative approach is to consider $\CaE_n$ as a function of $n$. 
If $\kappa>0$ then $\CaE_n$ is a strictly increasing function but if $\kappa<0$ then $\CaE_n$ is only increasing for small values of $n$, it has a maximun at the point $N_\kp = (1/(2|\kp|))(1-|\kp|(1+g))$ and then it becomes decreasing. Only the values of  $n$ 
placed in the incresing section of the curve determine normalizable wave functions. 

Summarizing, we have the following situation. 

\begin{enumerate}

\item[(a)]   Negative $\kappa<0$ case. There is a  finite number of bound states   $\Psi_{n}(x,g,\kp)$, with $n=0,1,2,\dots, n_\kp$ ($n_\kp$ denotes the greatest integer lower than $N_\kp$ ), and   the spectrum is bounded, not equidistant  
and with a gap between every two levels that decreases with $n$
$$
  E_0<E_1<E_2<\dots<E_{n_\kp} , 
   {\qquad} E_{n+1} - E_n = 2 \Bigl(1 - |\kp|\,\bigl(2n+2+g\bigr)\Bigr)\,(\hbar\,\al)\,.
$$

\item[(b)] Positive $\kappa>0$ case. There is an   infinite set of bound states
$\Psi_{n}(x,g,\kp)$, with $n=0,1,2,\dots$ and  the spectrum is unbounded, not equidistant and with a gap between every two consecutive levels
that increases with $n$
$$
  E_0<E_1<E_2<\dots<E_n<E_{n+1}<\dots  
   {\qquad} E_{n+1} - E_n = 2 \Bigl(1 + \kp\,\bigl(2n+2+g\bigr)\Bigr)\,(\hbar\,\al)\,.  
$$
\end{enumerate}

\section{Final Comments}

It is well known that the number of quantum Schr¬odinger solvable potentials is rather small mainly because quantum exactly solvability is a very fragile property. In fact, in most of cases the addition of a small perturbation to the potential (or the introduction of a small deformation in the Hamiltonian) breaks down the exact integrability. Thus, the exact solvability of the $(\kappa, g)$-dependent Hamiltonian $H (x, p_x ; \kappa, g)$ must be considered in fact as a very interesting property.

We conclude with the following two comments: 
First, it was proved in \cite{CaRS04}, \cite{CaRS07a} that the original  $\kappa$-dependent nonlinear system  (that is, the Hamiltonian  (\ref{HClasico}) but without the isotonic term) can be studied by making use of the Schr\"odinger factorization approach (ladder operators, intertwined Hamiltonians and shape-invariance property), therefore it seems natural that this new more general system (with $k_g\ne 0$) can also  be studied by this approach. 
Second, the nonlinear $\kappa$-dependet system can considered as a model of the harmonic oscillator on spaces of constant curvature (the parameter $\kappa$ represents the curvature of the space) \cite{CaRS07b}-\cite{CaRSJpa12}. 
According to this interpretation the quantum  isotonic oscillator can be correctly defined on the spaces on  $S_{\kp}^2$ ($\kappa>0$) and $H_{\kp}^2$ ($\kappa<0$). These two points are two interesting open questions  deserving to be studied.

\section*{Acknowledgments}

This work was supported by the research projects MTM--2012--33575 (MICINN, Madrid)  and DGA-E24/1 (DGA, Zaragoza). 

{\small

 }

\vfill\eject
\section*{\bf Figures}

\begin{figure}
\centerline{
\includegraphics{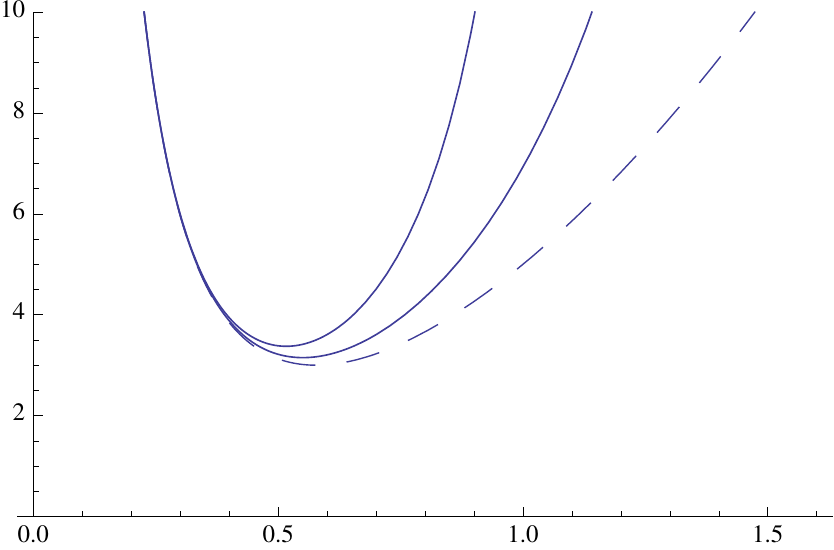}  } 

\caption{Plot of the potential $V_{\kp,g}(x)$ as a function of $x$ ($x>0$)
for $k_g=1$ and $\kappa=0$ (dash line) and some  positive values of $\kappa$.}
\label{Fig1}

{\vskip10pt}
\centerline{
\includegraphics{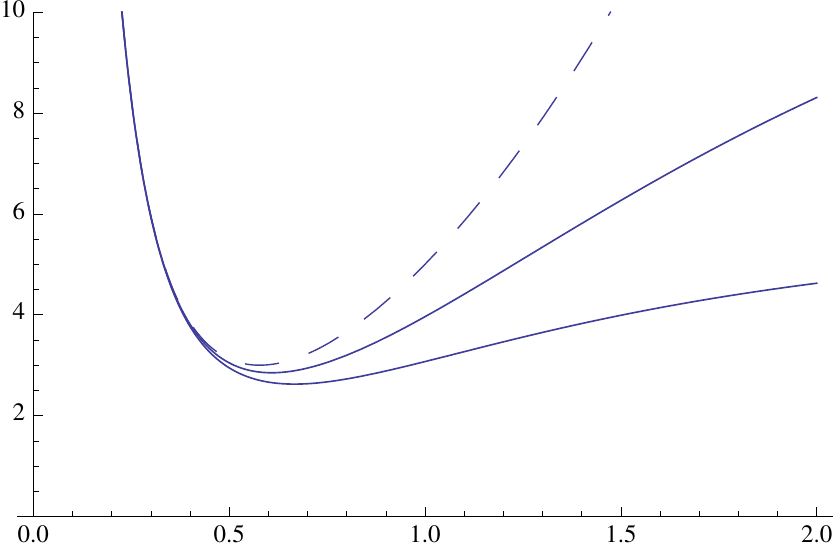}  } 

\caption{Plot of the potential $V_{\kp,g}(x)$ as a function of $x$ ($x>0$)
for $k_g=1$ and $\kappa=0$ (dash line) and some negative values of $\kappa$.}
\label{Fig2}
\end{figure}

\end{document}